# Coupler Conditioning and High Power Testing of ADS Spoke Cavity

CHEN Xu(陈旭)[1;2] MENG Fanbo(孟繁博) [1;2] MA Qiang(马强)[2] HUANG Tongming(黄彤明)[2] LIN Haiying(林海英)[2] PAN Weimin(潘卫民)[2]

1 (Graduate University of the Chinese Academy of Sciences, Beijing 100049, China)
2 (Institute of High Energy Physics, CAS, Beijing 100049, China)

**Abstract**：Power couplers, used in China-ADS proton linac injector I, are required to transfer 6kW RF power to the superconducting Spoke cavities. At present, first two couplers of coaxial design have been fabricated and accomplished high power test at IHEP. The test results indicated that couplers of this design are qualified to deliver 10kW RF power in continuous travelling wave mode. This paper described the coupler's room temperature test procedures and results, discussed the original high power test terminated due to serious out-gassing and after some modifications. In the final test, the couplers smoothly exceeded the design power level.

## 1. Introduction

A RF coupler is one of the key components in a superconducting (SC) proton linac. Coupler is a guarantee of impedance matching between RF source and load and its primary function is transmitting power from RF source to the beam loaded cavity efficiently. Coupler also provides vacuum barrier and holds a temperature gradient to the SC cavity. Its power handling capability and operation stability have to be tested prior to attach to the cryomodule. Two sets of coaxial couplers with planar ceramic window have been manufactured and accomplished processing and high power test at room temperature in late January. This type of coupler will be applied in China Accelerator Driven Sub-critical System (ADS) proton linac for beta=0.12 SC Spoke cavities which required to withstand 6kW continuous wave (CW) radio frequency (RF) power. Main parameters of this coupler are summarized in table 1.

Table1: main parameters of Spoke cavity couplers

| | |
|---|---|
| frequency | 325MHz |
| type | coaxial, antenna E-coupling |
| window | single, warm , coaxial disk |
| coupling | fixed |
| $Q_{ext}$ | 7.1E5 |
| input power | Max. 6kW |
| impedance | 50Ω |

## 2. Test stand

Two opposing couplers were attached to a test stand that allowed baking and following conditioning of both couplers simultaneously. The test stand was custom-designed, aimed to test the power transferring capability of the coupler[1]. Vacuum was formed between the two warm windows and pumped by an ion pump of 200L/s, in addition an aspirator pump of 400L/s. The test stand was supplied with up to 10kW RF power from a solid-state amplifier. The two couplers and the connecting test stand were matched to maximize the transmission RF power. The power was transferred from upstream coupler (1#) to the downstream coupler (2#) via the test stand and terminated with a matched water load at the end. A picture of block diagram for coupler test was shown in fig.1. The inner conductor and inner window frame shared common water cooling while the outer conductor was cooled by static air. Three monitoring ports were set close to the window including vacuum gauge, arc discharge and electron current monitor. These monitoring instruments are very important to prevent a fatal discharge breakdown of the ceramic windows[2]. One of the above signals exceeding the preset threshold will cause a fast cut off of power. Another vacuum gauge was placed at the side-wall of the test stand.

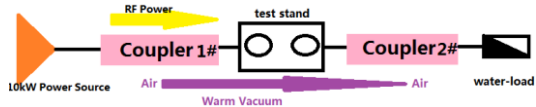

Fig.1: block diagram for coupler test

## 3. Baking

Prior to assembly, all coupler components and other associated parts were cleaned following our procedure to assure ultra cleanness[3]. Couplers were stored in special container for protection of oxidizing and transferred to the test stand. Handling with no linty rubber gloves only. For baking, the areas contains of windows, outer conductors, test stand wall, ion pump and aspirator pump using three insulated tape heating elements. Several pt100$_s$ were applied to the above areas to measure temperature changes. Baking temperature was controlled by a transformer below 115℃ in concern of the melting point of indium wire. We baked the system under vacuum for about 3 days obtaining a final vacuum pressure of $1.1 \times 10^{-5}$Pa, and the helium leak of the system was better than $9.5 \times 10^{-10}$mbarl/s. The baking process was illustrated in fig.2

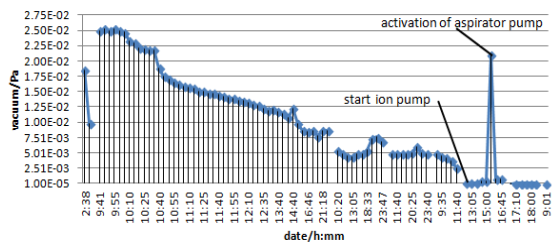

Fig. 2: record of baking process

## 4. Conditioning and high power test

### 4.1 The final test and observed phenomenon

Conditioning with high RF power before attaching the assembly to the clean cavity is important[4]. A photo of the power test site was shown in fig.3. At low RF power, we adjusted the two short circuits of our couplers to minimize reflected power in the transmission line. The voltage standing wave ratio (VSWR) value was adjusted to 1.15 at 325MHz that indicated less than 0.5% input power was reflected. This showed a good agreement with the previous simulation of S11 = -49dB.

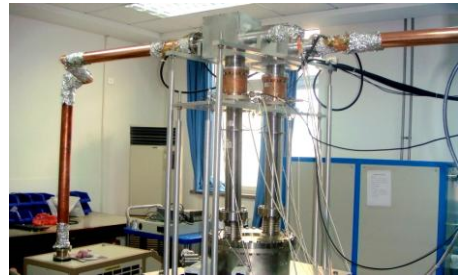

Fig. 3: photo of the test site

While increasing the RF power in short steps, a fast interlock system run which will immediately cut off the RF power if vacuum bursts or arc events occurred. Only conditioning with continuous wave method was adopted. It took about 8hrs of RF conditioning without major arc actions in the coupler to reach 10kW. History of high power test is shown in fig.4. No seriously vacuum bursts or discharging encountered. However, most of the interlock events occurred below 2.5kw and above 8.2kW. Between 2.5 and 8.2kW there were very few events and the conditioning time was less than three hours (see fig.5). Temperature around ceramic windows ranged from 31.1 to 31.9℃ at the power of 10kW. Fig.6 presented plot of RF power and temperature during the test.

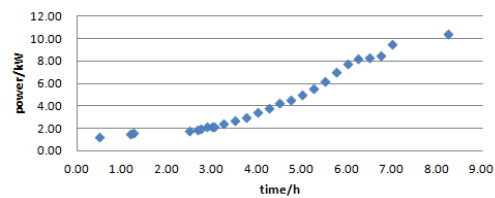

Fig. 4: history of high power test

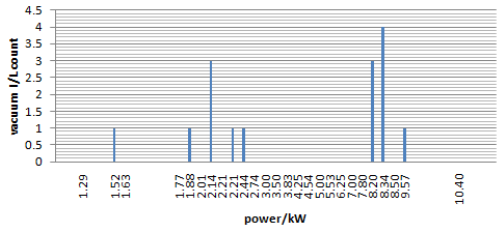

Fig. 5: number of vacuum interlocks

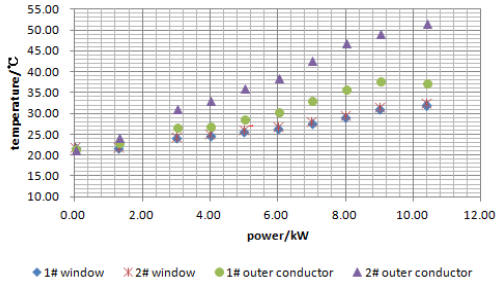

Fig. 6: temperature as a function of RF heating

In the process of testing, we observed unexpected temperature difference between the two outer conductors, one had a temperature of 6-10℃ higher than the other. (see fig.6). After exchanging one with the other, the same result as before. We supposed it was the inner coating surface that may lead to this phenomenon.

**4.2 Comparison with the original test**

In our original test, however, the couplers showed excessive out-gassing in some points that lead to several badly vacuum bursts and ion pump breakdown. Recovery of the vacuum pressure was very difficult and time-consuming. Pulsed power conditioning did not help. Therefore the strong vacuum burst forced us to terminate at 3.8kW.

We opened the test stand and remade the suspicious part where the tungsten inert gas welding (TIG) joints of the stainless steel parts seemed to cause out-gassing. TIG welding points may form an inside microscopic pores which was hardly to be detected in helium leak check and possibly lead to bad vacuum in power test. Furthermore, we reassembled the two couplers and baked them for a longer time for about 3 days with higher temperature than before did. Expanding baking time as well as raising baking temperature improved the efficiency of gas desorption and helped to shorten the processing time. With a series of checking and modifications, in our final RF test mentioned above, a maximum RF power was reached smoothly.

**5. Summary**

Our first two couplers have successfully passed RF levels of 10kW in excess of the design specifications. The ultimate power is limited by the klystron available. The original test revealed weak points of the fabrication, which has been improved in the final test. The test stand works well after some changes but the possible reasons related to out-gassing need to be better investigated. The long time high temperature baking was effective in desorbing gas and accelerating the processing test. Besides, TIG welding point properties related with vacuum must be controlled to avoid pits and unevenness. Temperature differences between the two outer conductors need to be well understood. In future test of other couplers, we hope to gain a better understanding of the processing limitations.